# Quantum annealing-based computed tomography using variational approach for a real-number image reconstruction

Akihiro Haga,

*Graduate School of Biomedical Sciences, Tokushima University, Tokushima 770-8503, Japan*
Electronic mail: haga@tokushima-u.ac.jp

## Abstract

Objective: Despite recent advancements in quantum computing, the limited number of available qubits has hindered progress in CT reconstruction. This study investigates the feasibility of utilizing quantum annealing-based computed tomography (QACT) with current quantum bit levels.

Approach: The QACT algorithm aims to precisely solve quadratic unconstrained binary optimization (QUBO) problems. Furthermore, a novel approach is proposed to reconstruct images by approximating real numbers using the variational method. This approach allows for accurate CT image reconstruction using a small number of qubits. The study examines the impact of projection data quantity and noise on various image sizes ranging from 4×4 to 24×24 pixels. The reconstructed results are compared against conventional reconstruction algorithms, namely maximum likelihood expectation maximization (MLEM) and filtered back projection (FBP).

Main result: By employing the variational approach and utilizing two qubits for each pixel of the image, accurate reconstruction was achieved with an adequate number of projections. Under conditions of abundant projections and lower noise levels, the image quality in QACT outperformed that of MLEM and FBP. However, in situations with limited projection data and in the presence of noise, the image quality in QACT was inferior to that in MLEM.

Significance: This study developed the QACT reconstruction algorithm using the variational approach for real-number reconstruction. Remarkably, only 2 qubits were required for each pixel representation, demonstrating their sufficiency for accurate reconstruction.

# Introduction

In recent years, quantum computing has rapidly become a growing field of research and development, marked by many significant breakthroughs and advances [1-5]. The development of new algorithms, hardware, and software tools has expanded the capabilities of quantum computers (QC) and opened up new possibilities for solving complex problems that are beyond the reach of classical computers.

One area where quantum computing shows great potential is in medical imaging [6], particularly in computed tomography (CT) image reconstruction [7-10]. CT imaging is a widely used diagnostic tool that relies on advanced computational techniques to generate detailed images of the human body. However, there are still significant challenges that need to be overcome to fully realize the potential of quantum computing for CT image reconstruction. One major challenge is the need to reduce radiation exposure associated with CT imaging while still maintaining image quality and diagnostic accuracy [11-14]. This is particularly important for patients who require repeated imaging studies, as cumulative radiation exposure can increase the risk of cancer and other health problems. In this context, iterative reconstruction approaches have shown promising results, although the process can be time-consuming and resource-intensive [14].

Recent developments in quantum computing offer the possibility of solving some of the challenges associated with CT image reconstruction. For example, quantum computers can perform certain types of mathematical calculations much faster than classical computers, which could significantly reduce the time required for image reconstruction. Additionally, quantum computing can potentially enable the use of more advanced image reconstruction algorithms that are not feasible on classical computers. There are two main approaches to quantum computing: the gate model [15] and the annealing model [16]. While most of the research in quantum computing has focused on the gate model, the annealing model has also shown promise for solving certain types of optimization problems [17]. In particular, the annealing model has been applied to image reconstruction in previous studies, but these studies have been limited to binary reconstruction or used only a small number of qubits to represent an integer number [8]. This limitation arises from the reliance on Quadratic Unconstrained Binary Optimization (QUBO), which is designed to optimize binary variables exclusively. Attempting a direct expansion of QUBO in computed tomography (CT) reconstruction with real numbers would necessitate a substantial number of qubits (an $n \times n$ image would require $n^2 q_{max}$-qubits with $q_{max}$ representing the maximum number of representations in each pixel). Consequently, the application of the QUBO problem for real CT reconstruction along this line becomes impractical.

In this paper, the potential of quantum computing for CT image reconstruction is explored and new methods that address some of the challenges associated with this process is proposed. Specifically, a novel approach to image reconstruction using the variational approach to approximate a real number [18,19], which allows for the accurate reconstruction of CT images using a small number of qubits, is proposed. The effectiveness of the proposed method is demonstrated through numerical simulations using both synthetic and real-world CT datasets.

Overall, this research emphasizes the potential of quantum computing in overcoming computational hurdles in CT image reconstruction. Through the development of tailored algorithms and techniques for quantum hardware, significant advancements can be achieved in the field. These advancements will lead to faster and more precise CT imaging, benefiting both patients and healthcare providers in the future.

## Methods

### Optimization problem using variational approach in QACT

In this section, the method of QACT using a variational approach for real numbers is described. Annealing is a technique used to solve optimization problems by encoding them into the global minimum of a given Hamiltonian,

$$H(\boldsymbol{\sigma}) = -\sum_{i,j} J_{ij}\sigma_i\sigma_j - h\sum_i \sigma_i, \quad (1)$$

as a QUBO [20], where $\sigma_i$ $(i \in \{0, \cdots, n-1\})$ is binary variables. The CT reconstruction problem can be expressed using equation (1) with the Hamiltonian given by,

$$H(\boldsymbol{\sigma}) = \sum_i (y_i^* - y_i)^2, \quad (2)$$

where the estimated and actual projections at the $i$th detector are denoted by $y_i^*$ and $y_i$, respectively. The projection is evaluated from the line-attenuation coefficient, $x_j^*$, as,

$$y_i^* = \sum_j A_{ij}x_j^*, \quad (3)$$

through the system matrix $A_{ij}$, which means the pass length in $j$th pixel for the $i$th projection. The CT reconstruction can be performed by minimizing equation (2) with respect to the variable parameters $x^* \in \mathbb{R}^{n \times n}$, where $n \times n$ is the reconstruction image size. In this study, $x^*$ in each pixel $j$ is expressed as,

$$x_j^* = 2^{-k_j} \sum_{q=0}^{q_{max}-1} 2^q \sigma_{q,j} + d_j, \quad (4)$$

where $d_j$ and $k_j$ are the parameters to express a real number with the binary expansion using $q_{max}$-qubits. In principle, a large number of qubits is required to obtain a good approximation of the real number. However, the available number of qubits is limited, especially since $n \times n$ image reconstruction requires $n^2 q_{max}$-qubits. Therefore, a variational approach was employed to achieve this; $d_j$ and $k_j$ in equation (4) are refined using,

$$d_j^{(k+1)} = x_j^{*(k)} - 2^{q_{max}-k_j^{(k+1)}-1}, \quad (5)$$

$$k_j^{(k+1)} = k_j^{(k)} + c_j, \quad (6)$$

until convergence. In this approach, the binary expansion of the real number is used to represent the pixel values using qubits, and the optimization is performed iteratively to achieve the best approximation with the limited number of available qubits.

The pseudocode 1 outlines the iterative calculation process used in this approach. The python library, *pyQUBO*, is used to map higher order polynomials to quadratic ones using the reduction by substitution method for the QACT reconstruction [21]. In this paper, the results with "*SimulatedAnnealingSampler*" sampler using a classical computer will be provided in the result section. I note that the same code was executable on the D-Wave 2000Q system by using "*AutoEmbeddingComposite*" sampler [22]. The parameters used for the iterative calculation and the library are set to $c_j = 0.5$, $k_j^{(0)} = 1$, and $q_{max} = 2$ in this study, and all results presented in the previous section of QACT are produced at 30 iterations using these parameters. With these specified values, the eigenvalues of the Hamilton (Eq. (2)) can be determined through time evolution adiabatically using quantum annealing. This is possible if the system, described by the Hamiltonian

$$H(s) = (1-s)H_B + sH_p, \quad (7)$$

maintains a sufficient energy gap between the first excited state and the ground state during time evolution from $s = 0$ to $s = 1$. Here $H_B$ and $H_p$ represent the initial and final (or problem) Hamiltonians, respectively. In accordance with Ref. [23], I assumed that $H_B = \sum \frac{1}{2}(1 - \sigma_x^{(i)})$, where $\sigma_x$ is the $x$-component of the Pauli matrix given by $\begin{pmatrix} 0 & 1 \\ 1 & 0 \end{pmatrix}$. As $H_p$ takes the form of Eq. (2), the eigenvalues of Eq. (7) were estimated as a function of time evolution index $s$ to confirm that the energy gap is sufficiently large during the iterative calculation in variational approach.

---

**Algorithm 1: A real-number reconstruction in QACT**

---

**input:** $A \in \mathbb{R}^{m \times n^2}$: System matrix,   $y \in \mathbb{R}^m$: Projection

**output:** $x \in \mathbb{R}^{n^2}$: Reconstructed image

**initialize:** $k \in \mathbb{R}^{n^2}$,  $d \in \mathbb{R}^{n^2}$

**for:**

       **initialize:** $\sigma$: $q_{max} n^2$-dimensional binary variables

       $x \leftarrow 2^{-k} \sum_q 2^q \sigma_q + d$:

       $H \leftarrow (Ax - y)^2$: Hamiltonian

       $\sigma \leftarrow$ SimulatedAnnealingSampler or AutoEmbeddingComposite

       $k \leftarrow k + c$:

       $d \leftarrow x - 2^{q_{max}-k-1}$:

**endfor:**

---

## Data preparation

Here, the preparation of the projection data used in this study is described. The CT geometry assumed in this study is shown in Fig. 1, which was constructed by referencing a realistic fan-beam CT machine (Activion16, Canon Medical System) [24]. All objects have a square format with 25.6 cm on each side, and are divided into 4×4 (6.4-cm pixel scale), 8×8 (3.2-cm pixel scale), 16×16 (1.6-cm pixel scale), and 24×24 (1.067-cm pixel scale) pixel grids. For the 4×4 object, the values of 0.3, 0.4, 0.8, and 0.2 was set in the middle 4 pixels and zero in the rest of the pixels. The Shepp-Logan phantom was used for the other image-size objects, which was imported into the Python code using the *phantominator* library with the desired size [25]. For the 24×24 object, more complex phantoms were also employed to confirm the validity of my reconstruction algorithm on images containing various real numbers. This phantom was created by downsampling from a real CT image [26] and then normalized to have a range of [0, 1]. I note that in solving the Quadratic Unconstrained Binary Optimization (QUBO) problem for image reconstruction, a significant amount of memory is required, even when utilizing classical computers Additionally, existing QCs such as D-Wave, which is currently available for QA, impose limitations on the number of qubits (typically around 2,000) [22]. Experiments could be possible with images of a certain size (ex. 30×30), but in this study, the image size was restricted to 24×24.

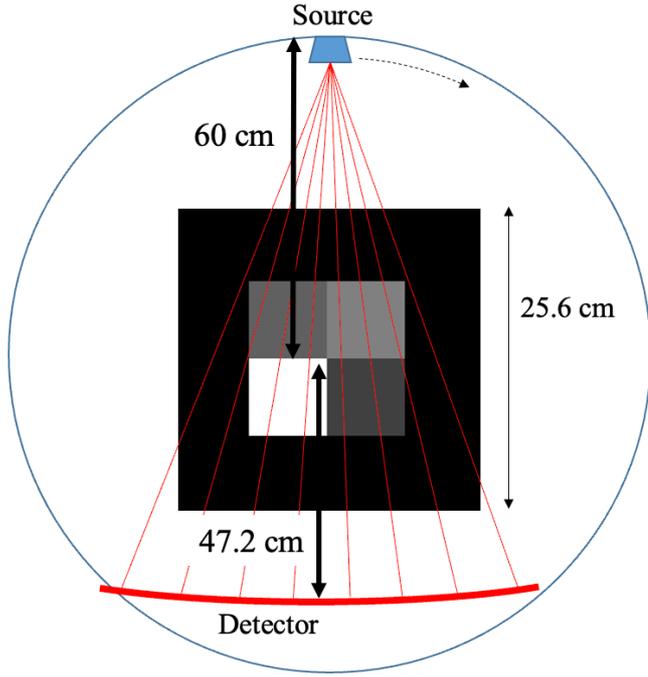

**Fig. 1 Geometry of the simulated CT system.**
In this study, an source-to-detector distance of 107.2 cm and isocenter-to-detector distance of 47.2 cm were assumed. A number of detector elements was set to the double of the image size ($2n$ for $n \times n$ image*)* .

The projection data was collected with $2n$ devices for $n \times n$ objects, aligned in an equally angle-spaced format where the distance between devices was set as 80% of the pixel scale in the object. Then, a 360-degree gantry rotation was performed. The projection was collected at 120-degree, 40-degree, 20-degree, and 10-degree intervals, corresponding to 3, 9, 18, and 36 gantry angle samplings, respectively.

An in-house ray-tracing model was used in the projection generation, as well as the corresponding system matrix generation. The base algorithm can be found at the Github indicated in Ref. [27].

**Noise model**

In this study, the projection data including noise is produced by sampling from following Poisson distribution,

$$p(I = k) = \frac{(I_i)^k e^{-I_i}}{k!}, \quad (6)$$

where $I_i = I_0 e^{-y_i}$ is the X-ray intensity generated in the $i$th detector of a virtual CT as shown in Fig. 1. The amount of the noise is controlled by $I_0$, and in this study, it was set to $I_0 \in \{10^1, 10^2, 10^3, 10^4, 10^5, 10^6\}$, where typically $10^5 < I_0 < 10^6$ provides the same order of signal-to-noise (SNR) ratio as that observed in a real CT system [23].

**Conventional models for comparison with QACT**

The reconstructed results obtained using the QACT algorithm were compared with those obtained using two representative conventional reconstruction algorithms, namely FBP [28] and MLEM [29]. In FBP, the Shepp-Logan filter [30] was used as the reconstruction filter. In MLEM, the identically independent Poisson distribution model was employed for the likelihood function, leading to the following iterative form [14]:

$$x_j^{*(k+1)} = x_j^{*(k)} \frac{\sum_i A_{ij} e^{-y_i^{*(k)}}}{\sum_i A_{ij} e^{-y_i}}, \quad (7)$$

where the maximum iteration number $k$ was set to 400 in all MLEM reconstruction scenarios, and the reconstructed image was then selected based on the root mean squared error (RMSE) and the structural similarity (SSIM) between the estimated projection $y_i^{*(k)}$ at the $k$th iteration and the projection generated from the ground truth image, $y_i$. An initial value of $x_j^{*(0)} = 0.1$ was set for all pixels.

## Results

### Reconstructed image

Figures 2-5 show reconstructed images of objects with various sizes (4×4, 8×8, 16×16, and 24×24) generated from the ground truth (GT) projections, as described in the Method section. The images in Figs. 2-5 were reconstructed with 36 (10-degree interval), 18 (20-degree intervals), 9 (40-degree intervals), and 3 (120-degree intervals) projection angles, respectively. Quantum annealing-based computed tomography images reconstructed using the proposed algorithm are denoted as QACT, while maximum likelihood expectation maximization (MLEM) and filtered back projection (FBP) are representative conventional approaches for CT reconstruction and are used for comparison purposes. From these figures, it can be observed that QACT outperforms FBP and is comparable to MLEM in terms of image quality with a larger number of projection angles, suggesting that QACT is useful in finding the exact solution for determined problems. It is emphasized that all images were reconstructed using only 2 qubits in each pixel, and the variational approach accurately reconstructed the images with adequate projection data (Fig. 1). The GT image in the bottom row of these figures contains various real numbers pixel by pixel, and the reconstructed image demonstrates the effectiveness of the proposed algorithm. However, image degradation is observed with a smaller number of projection angles (Figs. 3-5), implying that the use of QACT is inappropriate for ill-posed problems. Nevertheless, it is noted that this is not caused by QACT itself, but rather by the loss function to be minimized, which will be discussed later.

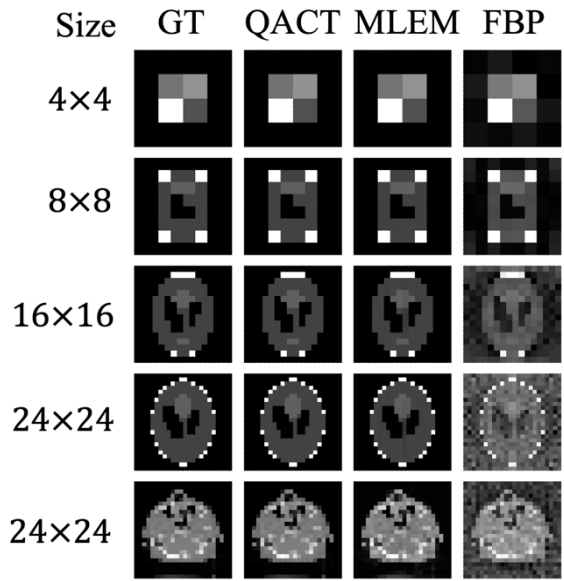

**Fig. 2 A real-number image reconstruction using 36 projection angles.** The top row is a 4×4 object with values of 0.3, 0.4, 0.8, and 0.2 in the middle 4 pixels and zero in the rest of the pixels. The 2nd -4th rows show the Shepp-Logan phantom. The bottom row is a downsampled image from a real CT image. GT: ground truth, QACT: quantum annealing based computed tomography, MLEM: maximum likelihood expectation maximization, and FBP: filtered back projection.

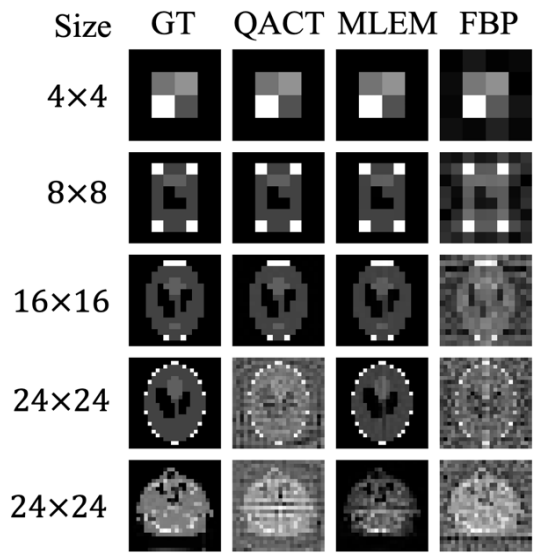

**Fig. 3 A real-number image reconstruction using 18 projection angles.** Same as in Fig. 2.

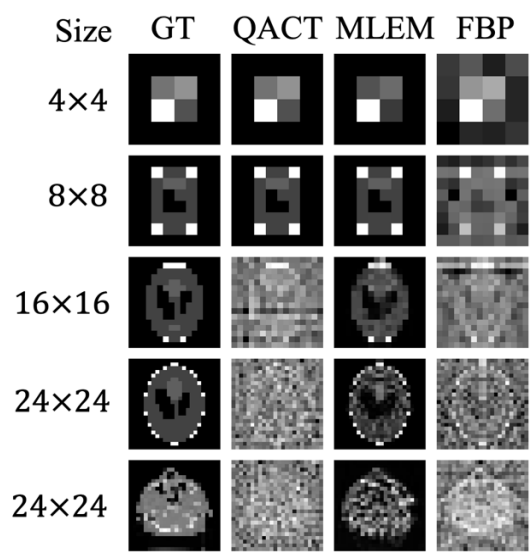

**Fig. 4 A real-number image reconstruction using 9 projection angles.** Same as in (Fig. 2).

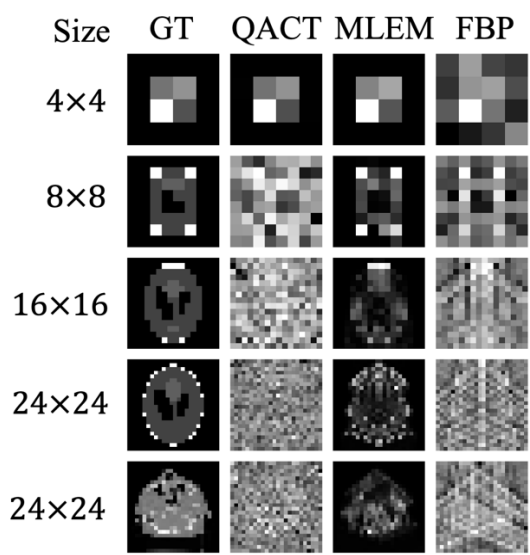

**Fig. 5 A real-number image reconstruction using 3 projection angles.** Same as in Fig. 2.

**Test on the number of projection angles**

Figure 6 shows the RMSE between the reconstructed images and their corresponding ground truth as a function of the number of projection angles. This figure quantitatively demonstrates that QACT outperforms for a larger number of projection angles while MLEM outperforms for a smaller number of projection angles. QACT can find an optimal solution with adequate number of projections.

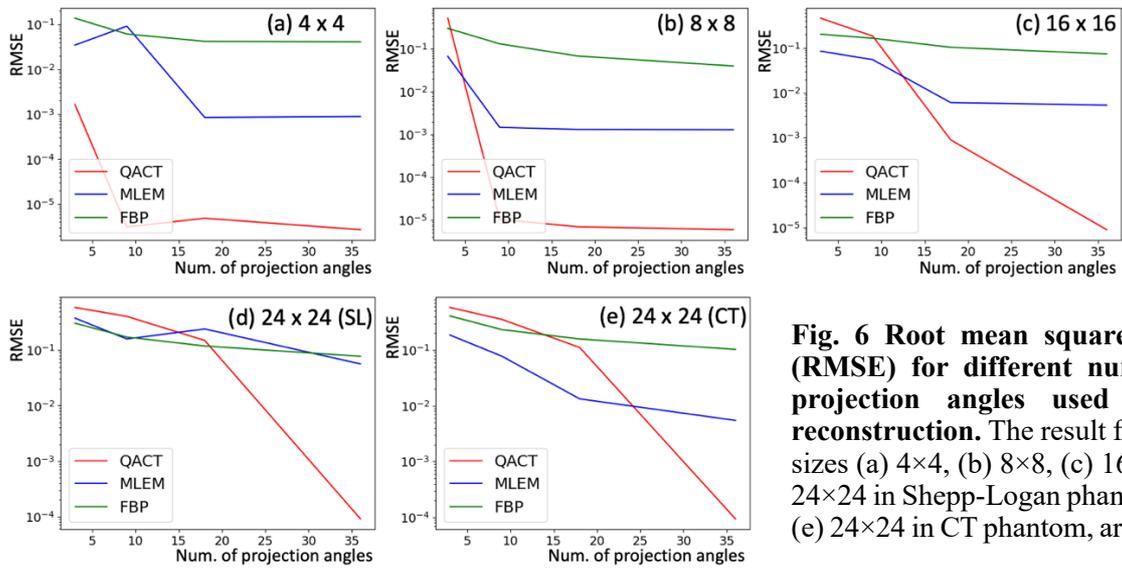

**Fig. 6 Root mean squared error (RMSE) for different number of projection angles used in the reconstruction.** The result for image sizes (a) 4×4, (b) 8×8, (c) 16×16, (d) 24×24 in Shepp-Logan phantom, and (e) 24×24 in CT phantom, are shown.

**Noise influence**

Noise is inevitable in actual CT acquisitions. In this study, the noise was modeled in the detectors using Poisson-distributed random noise for the image intensity. Details of the noise model are provided in the Method section. Figure 7 shows the effect of noise on the reconstructed image quality, where the RMSE (Fig. 7(a)) and SSIM (Fig. 7(b)) of a 16×16 pixel-size image are evaluated. Poisson noise with six different magnitudes is added to the corresponding projection images. From this figure, it is evident that the RMSE and SSIM change drastically with the amount of noise, indicating that QACT is sensitive to the magnitude of noise. In other words, QACT is less robust than MLEM and FBP when it comes to noise. However, the magnitude of the RMSE/SSIM in QACT is less than or comparable to that in MLEM and FBP for lower levels of noise.

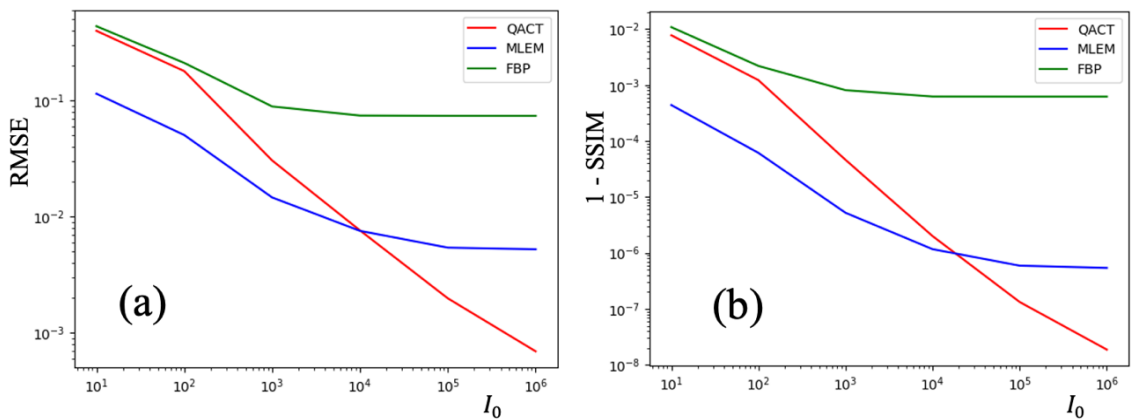

**Fig. 7 Image degradation caused by noise (a) RMSE and (b) SSIM.** Noise was added to the projections by assuming that the photon statistics in the X-ray detector follows a Poisson distribution. The horizontal axis indicates the X-ray intensity at the source ($I_0$). The added noise decreases as the horizontal axis increases. The reconstruction images using 36 projections for the 16×16 Shepp-Logan phantom are evaluated.

**Convergence in variational approach**

The algorithm used in this study for QACT is based on the variational method, which enables reconstruction of real number images with a small number of qubits (only 2 qubits in this study). As described in the Method section, this algorithm needs an iterative operation, and convergence provides useful information to verify the achievable accuracy in QACT reconstruction. Figure 8 depicts the convergence in reconstructed images with 36 projection angles, where the RMSE (Fig. 8(a)) and the SSIM (Fig. 8(b)) are shown as a function of the number of iterations for specific parameters used in this study (see equations (5) and (6) and the related sentences). For small pixel sizes, the accuracy improves with increasing iterations, while for larger pixel sizes, the accuracy seems to converge around 30 iterations. Although the accuracy is limited, with RMSE $\sim 10^{-4}$ in such cases, this would be satisfactory, because it is very lower than those in MLEM and FBP (Fig. 6(e)). The convergence feature in 24×24 (CT) (purple curve) suggests that the present QACT method can provide an arbitrary real number image even with only 2 qubits used. Thus, this method can be considered one of the most realistic approaches for medical applications of QACT. The variational approach can be analyzed based on the adiabatic process of the system Hamiltonian employed in the QA machine. Figure 9 illustrates the profiles of the energy gap between the lowest two eigenvalues in Eq. (7) as a function of the time index $s$. The results are presented for a 2×2 image (pixels removed with a value of zero from a 4×4 image) after 10, 15, 20, and 25 iterations. The energy gap of the initial Hamiltonian $H_B$ is one with its definition. As time progresses, the energy gap significantly increases in a small number of iterations, indicating that the adiabatic theorem works effectively. However, as seen at $s = 0.3$ in 25[th] iteration, the energy gap becomes relatively small. Although I cannot say that this behavior violates the adiabatic theorem, a diminished gap implies that the adiabatic process inadvertently yields a suboptimized result. This analysis suggests that the accuracy of the variational approach may have a limitation.

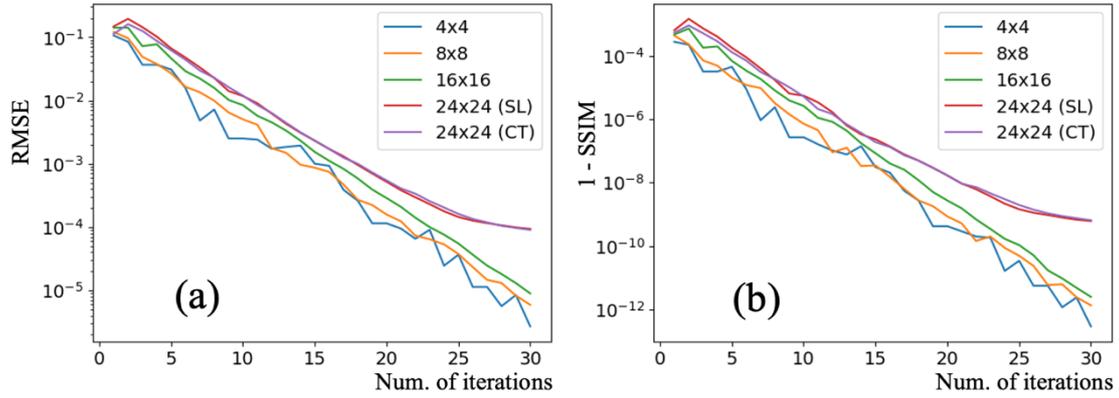

**Fig. 8 (a) Root mean squared error (RMSE) and (b) structural similarity (SSIM) at each iteration step in variational approach.** The reconstructed QACT images with 36 projection angles (Fig. 1) is used in this evaluation.

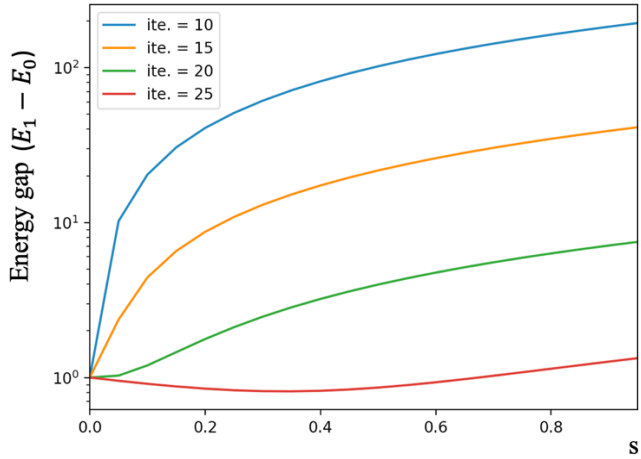

**Fig. 9 Energy gap simulation during time evolution in variational approach.** The energy gap between the lowest two states in Eq. (7) is shown as a function of time index s.

## Discussion

One significant challenge in the medical application of QC is the limited number of available qubits. Adiabatic QC produced by D-Wave has increased the number of qubits to be utilized, which has enhanced the field of QC [31-36]. Binary- or integer-based reconstruction has been proposed in recent years [8], but integer-based reconstruction still has difficulty reproducing quantitative images due to the shortage of qubits to represent pixel values even for 12-qubit representation, which is typically employed in a CT value. Thus, it may be difficult to develop a realistic QACT for direct medical applications. This study addressed this problem by using the variational approach to show that real-number based images can be reconstructed accurately even with only 2 qubits. In particular, the current approach reproduced a pixel-by-pixel different image, which would be impossible with integer-based reconstruction using binary expansion in a few qubits. In this context, the range dependent Hamiltonian algorithm has been proposed as a similar approach to the present method [36]. The subrange approach sets translation numbers, corresponding to $d_j$ in Eq. (4) of my method, but did not include $k_j$ and, more importantly, lacks the optimization process required for achieving highly accurate real-number simulations.

The present study also demonstrated that QACT can produce higher-quality images than conventional reconstruction algorithms with adequate projections. However, for 3 (120-degree interval) and 9 (40-degree interval) projection angles, MLEM outperformed QACT. Additionally, noise in QACT is more sensitive than that in MLEM. These findings may be due to the loss function used in QACT, which is the mean squared error (MSE) rather than the Poisson distribution used in MLEM to assume photon statistics. The MSE function is sensitive to statistical noise outliers [38, 39], and it is difficult to predict latent variables in undetermined problems such as image reconstruction for ultra-limited projection angles. The study used the MSE because it can be easily expanded as a QUBO problem [40] required by adiabatic QC using D-Wave. The use of other loss functions, such as weighted least squares [41, 42] including regularization terms [43, 44], could improve performance in the above situations without any change of the present algorithm. However, investigating the difference in loss function forms was outside the scope of this study. Rather, it is said that in a realistic noise presence and projection acquisition, QACT performance was satisfactory.

Finally, the study discussed the convergence in the variational approach to produce real-number images, where a part of the problem is run on QC and the post-processing is run on a classical computer. This approach converges in the convex-type problem used in this study. Convergence appeared at ~~20~~30 iterations with an RMSE of less than $10^{-4}$. The CT-number range in modern devices typically uses $2^{12}$ [45, 46], meaning the accuracy required is on the order of $10^{-4} \sim 10^{-3}$. Therefore, the choice of parameters, particularly the use of 2 qubits in each pixel representation, was appropriate within the scope of this study. However, larger pixel sizes may require different parameters to maintain image quality.

## Conclusion

This study developed the QACT reconstruction algorithm using a variational approach for real-number reconstruction. Only 2 qubits were used in each pixel representation, which showed to be sufficient for accurate reconstruction. The proposed algorithm demonstrated superior performance compared to conventional reconstruction algorithms such as MLEM and FBP, particularly in scenarios with abundant projections and lower noise levels. However, MLEM outperformed QACT in terms of image quality when fewer projections and/or higher noise conditions were present. This suggests that further development should focus on constructing the loss function as a QUBO problem or employing alternative approaches, as well as reducing the reliance on the number of quantum qubits.

**Acknowledgements**

The author thanks Yataro Horikawa (Juntendo University) and Yoshi-Hide Sato (Tokushima University) for fruitful discussions on quantum computing.

**Competing Interests**

The authors declare no competing interests.